\documentclass[aps,prl,twocolumn,showpacs,amssymb,amsfonts,longbibliography]{revtex4-1}
\usepackage{graphicx}
\usepackage{dcolumn} 
\usepackage{bm}      

\begin{document}

\title{Field Driven Quantum Criticality in the Spinel Magnet ZnCr$_2$Se$_4$}

\author{C. C. Gu$^1$}
\author{Z. Y. Zhao$^{2,3}$}
\author{X. L. Chen$^1$}
\author{M. Lee$^{4,5}$}
\author{E. S. Choi$^4$}
\author{Y. Y. Han$^1$}
\author{L. S. Ling$^1$}
\author{L. Pi$^{1,2,7}$}
\author{Y. H. Zhang$^{1,7}$}
\author{G. Chen$^{6,7}$}
\email{gangchen.physics@gmail.com}
\author{Z. R. Yang$^{1,7,8}$}
\email{zryang@issp.ac.cn}
\author{H. D. Zhou$^{9,10}$}
\email{hzhou10@utk.edu}
\author{X. F. Sun$^{2,7,8}$}
\email{xfsun@ustc.edu.cn}

\affiliation{$^1$Anhui Province Key Laboratory of Condensed Matter Physics at Extreme Conditions,
High Magnetic Field Laboratory, Chinese Academy of Sciences, Hefei, Anhui 230031, People's Republic of China}
\affiliation{$^2$Department of Physics, Hefei National Laboratory for Physical Sciences at Microscale, and Key Laboratory of Strongly-Coupled Quantum Matter Physics (CAS), University of Science and Technology of China, Hefei, Anhui 230026, People's Republic of China}
\affiliation{$^3$Fujian Institute of Research on the Structure of Matter, Chinese Academy of Sciences, Fuzhou, Fujian 350002, People's Republic of China}
\affiliation{$^4$National High Magnetic Field Laboratory,
Florida State University, Tallahassee, FL 32306-4005, USA}
\affiliation{$^5$Department of Physics, Florida State University,
Tallahassee, FL 32306-3016, USA}
\affiliation{$^6$State Key Laboratory of Surface Physics,
Center for Field Theory \& Particle Physics
Department of Physics, Fudan University, Shanghai, 200433, China}
\affiliation{$^7$Collaborative Innovation Center of Advanced Microstructures,
Nanjing, Jiangsu 210093, People's Republic of China}
\affiliation{$^8$Institute of Physical Science and Information Technology, Anhui University, Hefei, Anhui 230601, People's Republic of China}
\affiliation{$^9$Key laboratory of Artificial Structures and Quantum Control (Ministry of Education),
School of Physics and Astronomy, Shanghai JiaoTong University, Shanghai, 200240, China}
\affiliation{$^{10}$Department of Physics and Astronomy, University of Tennessee,
Knoxville, Tennessee 37996-1200, USA}
\date{\today}

\begin{abstract}
We report detailed dc and ac magnetic susceptibilities, specific heat, and
thermal conductivity measurements on the frustrated magnet ZnCr$_2$Se$_4$. At
low temperatures, with increasing magnetic field, this spinel material
goes through a series of spin state transitions from the helix spin state
to the spiral spin state and then to the fully polarized state. Our results
indicate a direct quantum phase transition from the spiral spin state to
the fully polarized state. As the system approaches the quantum criticality,
we find strong quantum fluctuations of the spins with the behaviors such as
an unconventional $T^2$-dependent specific heat and temperature independent mean
free path for the thermal transport. We complete the full phase diagram of
ZnCr$_2$Se$_4$ under the external magnetic field and propose the
possibility of frustrated quantum criticality with extended densities of
critical modes to account for the unusual low-energy excitations in the
vicinity of the criticality. Our results reveal that ZnCr$_2$Se$_4$ is a
rare example of 3D magnet exhibiting a field-driven quantum criticality
with unconventional properties.
\end{abstract}


\maketitle

Since the new centuary, quantum phase transition has emerged as an
important subject in modern condensed matter physics~\cite{Sachdevbook}.
Quantum phase transition and quantum criticality are associated with
qualitative but continuous changes in relevant physical properties of
the underlying quantum many-body system at absolute zero
temperature~\cite{RevModPhys.75.913,Sachdevbook}.
In the vicinity of quantum criticality, the low-energy and
long-distance properties are controlled by the quantum
fluctuation and the critical modes of the phase transition
such that certain interesting and universal scaling laws
could arise. It is well-known that quantum criticality often
occurs in the system with competing interactions where
different interactions favor distinct phases or orders.
Many physical systems such as the high-temperature superconducting
cuprates~\cite{RevModPhys.75.913}, heavy fermion and Kondo lattice
materials~\cite{Qimiao}, Fermi liquid metals with spin density
wave instability~\cite{RevModPhys.79.1015}, and Mott insulators
have been proposed to be realizations of quantum criticality~\cite{Sachdevbook}.
For superconductors and metals, the multiple low-energy
degrees of freedom and orders may complicate
the critical phenomena and the experimental interpretation.
In contrast, Mott insulators with large charge gaps are
primarily described by spin and/or orbital degrees of freedom
and may have the advantage of simplicity in revealing critical
behaviors.

The Ising magnets CoNb$_2$O$_6$ and LiHoF$_4$ in external
magnetic fields realize the quantum Ising model and
transition~\cite{Coldea2011,Liang2015,PhysRevX.4.031008,PhysRevLett.112.137403,PhysRevB.75.054426,PhysRevB.70.144411}.
External magnetic fields in dimerized magnets like
han purple BaCuSi$_2$O$_6$~\cite{PhysRevLett.93.087203,nphys893}
induce a triplon Bose-Einstein condensation transition.
In a more complicated example of the diamond lattice
antiferromagnet FeSc$_2$S$_4$~\cite{FeScS1, FeScS2, FeScS3, FeScS4, FeScS5}, it is the competition
between the superexchange interaction and the on-site
spin-orbital coupling that drives a quantum phase transition
from the antiferromagnetic order to the spin-orbital singlet
phase~\cite{PhysRevLett.102.096406,PhysRevB.80.224409}.
These known examples of quantum phase transitions in strong Mott
insulating materials with spin degrees of freedom are described
by simple Ising or Gaussian criticality where there are
 discrete number of critical modes governing the low-energy properties.
In this Letter, we explore the magnetic properties of a
three-dimensional frustrated magnetic material ZnCr$_2$Se$_4$.
From the thermodynamic, dynamic susceptibility and thermal transport
measurements, we demonstrate that there exists a field-driven quantum
criticality with unusual properties such as a $T^2$-dependent specific
heat and temperature independent mean free path for the thermal transport.
Our quantum criticality has extended numbers of critical modes
and is beyond the simple Ising or Gaussian
criticality among the existing materials that have been reported before.

In the spinel compound ZnCr$_2$Se$_4$, the Cr$^{3+}$ ion hosts the localized
electrons and give rise to the spin-3/2 (Cr$^{3+}$) local moments that form
a 3D pyrochlore lattice. The reported dielectric polarization~\cite{ZnSeS1},
magnetization and ultrasound~\cite{ZnSeS2}, neutron and synchrotron
x-ray~\cite{ZnSeS3, ZnSeS4} studies have shown that, with increasing
magnetic field, this system goes from helix spin state to spiral
spin state to an unidentified regime, and then fully polarized state
at the measured temperatures. Two possibilities have been proposed
for this unidentified regime, an umbrella state and a spin nematic
state~\cite{ZnSeS2,ZnCrO1}. Both the umbrella state and a spin nematic
state break the spin rotational symmetry. We address this unidentified
regime by completing the magnetic phase diagram of ZnCr$_2$Se$_4$ with
dc and ac susceptibility, specific heat, and thermal conductivity
measurements. We do not observe signatures of symmetry breaking in the
previously unidentified regime down to the lowest measured temperature.
We attribute our experimental results to a quantum critical point (QCP)
between the spiral spin state and the polarized state, and identify the
previously unidentified regime as the quantum critical regime.

\begin{figure}[tbp]
\linespread{1}
\par
\begin{center}
\includegraphics[width=3.0 in]{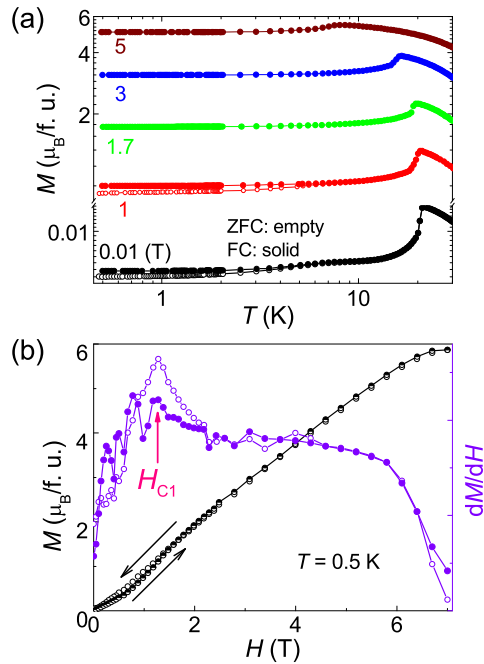}
\end{center}
\par
\caption{(Color online.)
(a) The temperature dependence of zero field cooling (ZFC)
and field cooling (FC) dc magnetizations at different applied fields.
(b) The dc magnetization measured at 0.5 K and its d\emph{M}/d\emph{H} curve.}
\label{fig1}
\end{figure}

The experimental details are listed in the online supplemental materials~\cite{suppl}. The dc magnetization measured at 0.01 T in Fig.~\ref{fig1}(a) shows a
pronounced peak at $T_{\text{N}} = 21$ K, corresponding to the
antiferromagnetic (AFM) order accompanied by a cubic to tetragonal
structural transition as previsouly reported~\cite{ZnSeS3}.
With increasing fields, the peak shifts to lower temperatures.
The dc magnetization measured at 0.5 K in Fig.~\ref{fig1}(b)
shows an anomaly near $H_{\text{C1}}\sim 1.6$ T,
which is more evident as a peak on the d\emph{M}/d$H$ curve.
As previous studies reported, the magnetic domain
reorientations occurs at this critical field $H_{\text{C1}}$
and above which, the helix spin structure is transformed into a
tilted conical one~\cite{ZnSeS1, ZnSeS2, ZnSeS3, ZnSeS4}. Due to
the reorientation of magnetic domains, the magnetization displays
hysteresis when the field is ramping down below $H_{\text{C1}}$.
This reorientation is also revealed as an irreversibility
between the ZFC and FC curves below 8 K for susceptibility
measured at 0.01 T, while it is suppressed completely at $H\geq1.7$ T.

\begin{figure}[tbp]
\linespread{1}
\par
\begin{center}
\includegraphics[width=3.0 in]{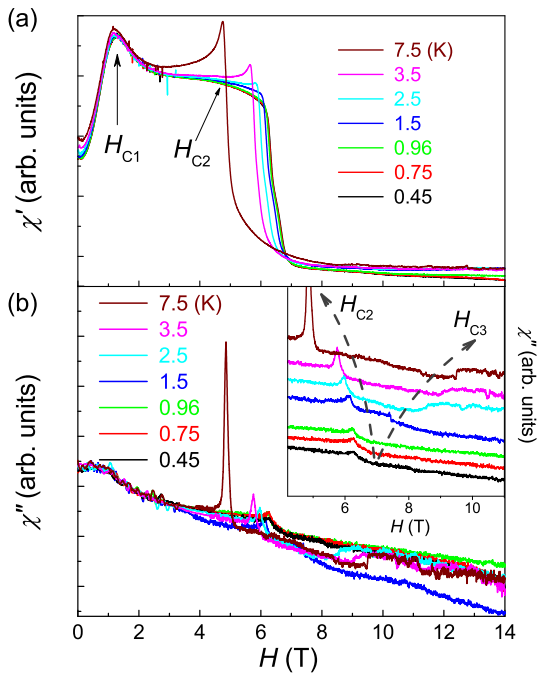}
\end{center}
\par
\caption{(Color online.) The magnetic field dependence of ac
susceptibility at several temperatures: (a) the real component;
(b) the imaginary component. The inset of (b) shows the zoom-in
of the high-field data. The arrows indicate the evolution of
high-field anomalies with increasing temperatures.}
\label{fig2}
\end{figure}

The real part of ac susceptibility $\chi^{\prime}$ in Fig.~\ref{fig2}(a)
clearly shows two peaks at $H_{\text{C1}}$ and $H_{\text{C2}}$.
Here, $H_{\text{C1}}$ is consistent with the $H_{\text{C1}}$ obtained
from the magnetization data above. $H_{\text{C2}}$ is consistent
with the reported $H_{\text{C2}}$ value, above which the spiral
spin structure is suppressed with a concomitant structure transition
from tetragonal to cubic. Meanwhile, a small bump at $H_{\text{C1}}$,
a sharp peak at $H_{\text{C2}}$, and a step-like anomaly near 9.5 T
are clearly seen for the imaginary part ($\chi^{\prime\prime}$)
measured at 7.5 K. This step-like anomaly is in accordance with the
plateau observed from the sound velocity measurements around 10 T at 2 K,
which has been correlated to the onset of fully polarized magnetic
phase at $H_{\text{C3}}$~\cite{ZnSeS2}. Upon further cooling,
$H_{\text{C3}}$ moves to lower fields and is hardly discernible
below 1.5 K from the ac susceptibility measurement, while
$H_{\text{C2}}$ shifts to higher fields (see the inset of
Fig.~\ref{fig2}(b)).

\begin{figure}[t]
\linespread{1}
\par
\begin{center}
\includegraphics[width=3.4 in]{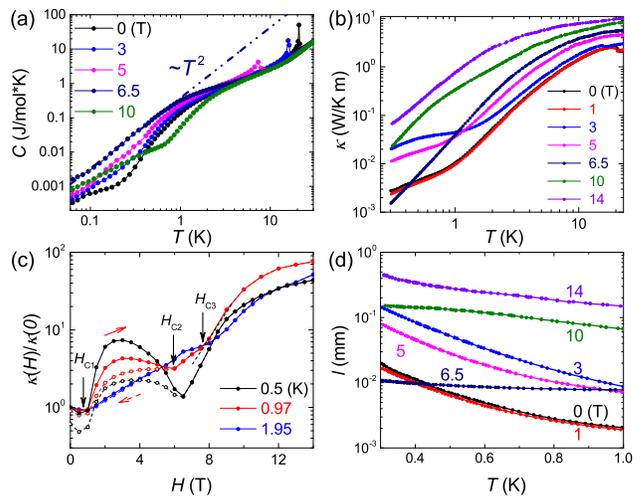}
\end{center}
\par
\caption{(Color online.)
(a) The temperature dependence of specific heat at several magnetic
fields from 0.06 K to 30 K. The dashed line represents the $T^{2}$ dependence.
(b) The temperature dependence of thermal conductivity from 0.3 K to 30 K at
various magnetic fields. (c) The field dependence of thermal conductivity
at selected temperatures below 2 K. (d) The calculated mean free path.}
\label{fig3}
\end{figure}

At zero magnetic field, the specific heat in Fig.~\ref{fig3}(a)
shows a sharp peak at $T_{\text{N}} = 21$ K, which shifts to
a lower temperature with increasing magnetic field and
disappears completely at 6.5 T. Moreover,
a small low-temperature hump around $1\sim 2$ K is observed at
zero magnetic field, which is enhanced with increasing field up to 6.5 T
and then strongly suppressed at 10 T. This kind of field dependence
is very different from the usual Schottky anomaly of magnetic
specific heat. Therefore, this anomaly could be originated from
the spin fluctuations. It is consistent with the recent
neutron-diffraction studies that reveal broad diffuse
scattering due to spin fluctuations in the long-range-ordered
state at temperatures down to 4 K~\cite{ZnSeS3, ZnSeS4}.
The strongest hump at 6.5 T suggests stronger spin fluctuations
around this field. Below 1 K, we tend to fit the heat capacity
data at 6.5 T with a $\gamma T^\alpha$ behavior. The obtained
result is $T^2$ down to the lowest temperature of 0.06 K. Here
we assume the lattice contribution of specific heat at so low
temperatures is negligible, and then the $T^2$ behavior for
6.5 T data is abnormal for a 3D magnet.

To further manifest the dynamic properties of the system under the magnetic
field, we carry out the thermal conductivity measurement.
As we depict in Fig.~\ref{fig3}(b), the thermal conductivity $\kappa$ at
0 T shows a structural-transition-related anomaly at $T_{\text{N}} = 21$ K
and a strong weakness of the $\kappa(T)$ slope around 1 K that
should be related to the spin fluctuations observed from specific
heat. With increasing magnetic field, $T_{\text{N}}$ shifts to
lower temperatures and disappears at $H \geq 5$ T. The slope change
around 1 K is not sensitive to the magnetic field but diminishes at
$H \geq 6.5$ T. While the $\kappa$ mainly shows a gradual increase
with increasing magnetic field at high temperatures ($T > 3$ K),
its field dependence is complicated at low temperatures ($T < 1$ K),
which is more clearly demonstrated in Fig.~\ref{fig3}(c).

At 1.95 K, the $\kappa(H)/\kappa(0)$ curve in Fig.~\ref{fig3}(c)
shows three weak anomalies at $\sim$ 1, 5.5 and 8 T, which
correspond to $H_{\text{C1}}$, $H_{\text{C2}}$, and $H_{\text{C3}}$,
respectively. At $H_{\text{C1}}$, a spin re-orientation appears,
which is related to a minimizing of the anisotropy gap and a
sudden increase of the AFM magnon excitations. This could
cause an enhancement of magnon scattering on phonons and
the low-field decrease of $\kappa$. The second anomaly
at $H_{\text{C2}}$, which becomes clearer at 0.97 K, is
demonstrated as a dip-like suppression of $\kappa$ and
is likely due to the spin fluctuations at $H_{\text{C2}}$.
The third anomaly at $H_{\text{C3}}$, identified as a
quicker increase of $\kappa$, is apparently due to the
strong suppression of spin fluctuations associated with
the transition or crossover from that unidentified regime to the
fully polarized spin state. The spin fluctuations are
strongly suppressed in the fully polarized spin state
because the spin excitation is gapped at low energies.
At lower temperatures that were not accessed in the
previous experiments~\cite{ZnSeS1,ZnSeS2,ZnSeS3,ZnSeS4},
the anomalies at $H_{\text{C2}}$ and $H_{\text{C3}}$
tend to merge, consistent with the opposite temperature
dependencies of these two critical fields observed
from our ac susceptibility measurement. In particular, at
0.5 K these two anomalies merge into a single one at 6.5 T
and the $\kappa(H)/\kappa(0)$ curve shows a deep
valley at the background of field-induced enhancement.
This is consistent with the specific heat result
showing that the spin fluctuation is the strongest
around 6.5 T. As we will explain in detail, both the
specific heat and the thermal transport results
suggest the existence of the quantum criticality at 6.5 T.

Before getting onto our intepretation, we here calculate the
phonon mean free path from $\kappa$ using a standard method~\cite{mfp}.
We choose the Debye temperature to be 308 K~\cite{Debye} and
assume $\kappa$ is primarily phononic. The results are depicted
in Fig.~\ref{fig3}(d). At 0 T, the phonon mean free path
($l\sim 10^{-2}$ mm) is nearly two orders of magnitude
smaller than the sample size ($\sim 1$ mm) even at the
lowest temperature of 0.3 K. This means that
the phonon scattering is still active at such low temperatures.
Since the phonon scatterings caused by phonons, impurities,
and other crystal defects are known to be quenched at low
temperatures, there must be some magnetic scattering
processes. Also because of the small mean free path $l$,
the magnetic excitations are not likely to make a sizable
contribution to the heat transport. With increasing magnetic
fields, $l$ is generally enhanced, indicating a suppression
of magnetic scatterings. Under the highest field of 14 T,
the phonon mean free path approaches the sample size,
which indicates the complete suppression of spin fluctuations
in the polarized state. This is consistent with the gapped
spin excitations for the fully polarized spin state.
In contract, at 6.5 T, $l$ drops back to 10$^{-2}$ mm
size with no obvious temperature dependence below 1 K.

\begin{figure}[tbp]
\linespread{1}
\par
\begin{center}
\includegraphics[width=3.2 in]{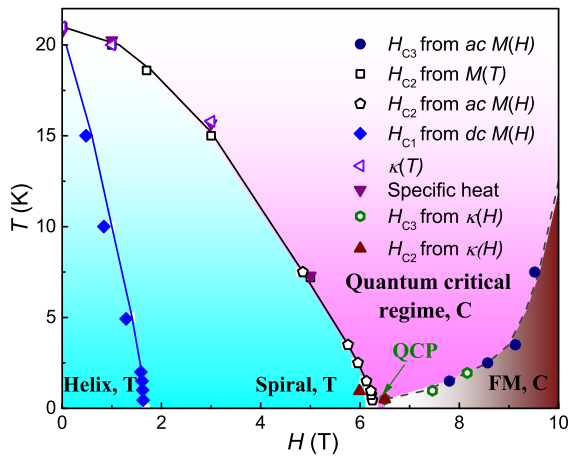}
\end{center}
\par
\caption{(Color online.) The $H$-$T$ phase diagram of ZnCr$_2$Se$_4$.
``T" and ``C" refer to the tetragonal and the cubic structure,
respectively. ``Helix'', ``Spiral'', ``FM'' stand for the
helix spin state, spiral spin state, and spin-fully polarized state,
respectively. A QCP is deduced between the spiral spin state and the
polarized phase. The solid (dashed) boundary refers to actual phase
transition (crossover). The pink region is marked as the quantum
critical regime. See the main text for the detailed discussion. }
\label{fig4}
\end{figure}

A detailed $H$-$T$ phase diagram of ZnCr$_2$Se$_4$ was constructed
in Fig.~\ref{fig4} by using the phase transition temperatures and
critical fields obtained from our above measurements. By comparing
to the reported phase diagram~\cite{ZnSeS2}, two important new features
were observed in this full phase diagram with lower temperatures
and higher magnetic fields. One is that the phase transition
temperature for the spiral spin structure is suppressed to zero
temperature with increasing fields before the system enters the
fully polarized state. Therefore, there is a direct quantum phase
transition between the spiral spin state and the polarized phase,
and this transition is marked as the QCP in Fig.~\ref{fig4}.
The other one is that the previous unidentified regime between the
spiral state and the fully polarized state does not persist down
to the lowest temperature. Note that our measurements were carried
out at a much lower temperature than the previous reports.
Thus, in Fig.~\ref{fig4} this previously unidentified regime is
naturally identified as the quantum critical regime that is the
finite temperature extention of the quantum criticality.

Why is the previously unidentified regime not an umbrella state or
a spin nematic state? As we have pointed out earlier, both states
break the spin rotational symmetry, and the former may break the
lattice translation. This is a 3D system, and this kind of symmetry
breaking should persist down to zero temperature and cover a finite
parameter regime. This finite-range phase is not observed at the lowest
temperature. For the same reason, the symmetry should be restored
at high enough temperatures via a phase transition. Such a thermodynamic
phase transition is clearly not observed in the heat capacity and
thermal transport measurements.

The spin spiral state and the fully polarized state are distinct phases
with different symmetry properties. The latter is translational invariant
and fully gapped, while the former breaks the lattice symmetry and
spin rotational symmetry. There must be a phase transition separating
them, and this quantum phase transition is manifested as the QCP
at 6.5 T in Fig.~\ref{fig4}. What is the property of this criticality?
The heat capacity was found to behave as $T^2$ at low temperatures at the QCP,
indicating much larger density of states than the simple Gaussian fixed point.
For a Gaussian fixed point, we would expect the heat capacity as $T^3$ up
to a logarithmic correction due to the critical fluctuations. The $T^2$
heat capacity suggests that the low-energy density of states
should scale as $D(\epsilon) \sim \epsilon$ with the energy $\epsilon$.
We know that the nodal line semimetal with symmetry and topologically
protected line degeneracies has this extended density of states
when the Fermi energy is tuned to the degenerate point~\cite{PhysRevB.84.235126}.
However, our system is purely bosonic with spin degrees of
freedom, and there is no emergent fermionic statistics.
To support $D(\epsilon) \sim \epsilon$ at the QCP, we would have the
critical modes to be degenerate or almost degenerate along the lines
in the reciprocal space such that the current thermodynamic measurement
cannot resolve them. It has been known that the frustrated
spin interactions could lead to such line degeneracies for the critical modes
and the resulting frustrated quantum criticality~\cite{PhysRevB.81.214419,PhysRevLett.109.016402}.
The possibility that infinite modes with line degeneracies become
critical at the same time is an unconventional feature of this QCP.
These critical modes scatter the phonon strongly and suppress the
thermal transport near the criticality.
It will be interesting to directly probe these degenerate
modes with inelastic neutron scattering and explore the
fates of the critical modes on both sides of the QCP.
Our thermal transport results also call for
further theoretical effects on the scattering between
the extended density of critical modes and the low-energy
phonons near the criticality.

Finally the system displays different lattice structures for different
magnetic phases in the phase diagram. Both the helix and the spiral spin
states have the tetragonal structure, while the quantum critical regime
and the fully polarized state have the cubic structure.
This is simply the consequence of the spin-lattice coupling.
The helix and the spiral spin states break the lattice cubic symmetry,
and this symmetry is transmitted to the lattice via the spin-lattice
coupling. The quantum critical regime
and the fully polarized state are uniform states and restore the lattice symmetry.
The correlation between the sound velocity and the magnetic structure
in the previous experiments has a similar origin~\cite{ZnSeS2}.

In summary, by completing the $H$-$T$ phase diagram of ZnCr$_2$Se$_4$,
we demonstrate the existence of QCP and quantum critical regime
induced by applied magnetic phase in this 3D magnet.
Our finding of the unconventional quantum criticality
calls for future works and is likely to provide an unique
example of frustrated quantum criticality for further studies.

\emph{Acknowledgments.}---This research was supported
by the National Key Research and Development Program of China
(Grant No. 2016YFA0401804), the National Natural Science Foundation
of China (Grant Nos. 11574323, U1632275) and the Natural Science
Foundation of Anhui Province (Grant No. 1708085QA19). X.F.S.
acknowledges support from the National Natural Science Foundation
of China (Grant Nos. 11374277 and U1532147), the National Basic
Research Program of China (Grant Nos. 2015CB921201 and 2016YFA0300103)
and the Innovative Program of Development Foundation of Hefei Center
for Physical Science and Technology. G.C. thanks for the support from
the ministry of science and technology of China with the grant
No. 2016YFA0301001, the initiative research funds and the program
of first-class University construction of Fudan University,
and the thousand-youth-talent program of China. H.D.Z. thanks
for the support from the Ministry of Science and Technology of
China with Grant No. 2016YFA0300500 and from NSF-DMR with
Grant No. NSF-DMR-1350002. Z.Y.Z. acknowledges support from
the National Natural Science Foundation of China (Grant No. 51702320).
M.L. and E.S.C. acknowledge the support from NSF-DMR-1309146.
The work at NHMFL is supported by NSF-DMR-1157490 and the
State of Florida. The x-ray work was performed at HPCAT (Sector 16), Advanced Photon Source,
Argonne National Laboratory. HPCAT operations are supported by DOE-NNSA under Award No. DE-NA0001974
and DOE-BES under Award No. DE-FG02-99ER45775, with partial instrumentation funding by NSF. The Advanced Photon
Source is a US Department of Energy (DOE) Office of Science User Facility operated for the DOE Office of Science by
Argonne National Laboratory under Contract No. DE-AC02-06CH11357.  

C.G. and Z.Y.Z. contributed equally to this work.

\end{document}


Supplemental Online Material

\title{Field Driven Quantum Criticality in the Spinel Magnet ZnCr$_2$Se$_4$}

\author{C. C. Gu$^1$}
\author{Z. Y. Zhao$^{2,3}$}
\author{X. L. Chen$^1$}
\author{M. Lee$^{4,5}$}
\author{E. S. Choi$^4$}
\author{Y. Y. Han$^1$}
\author{L. S. Ling$^1$}
\author{L. Pi$^{1,2,7}$}
\author{Y. H. Zhang$^{1,7}$}
\author{G. Chen$^{6,7}$}
\email{gangchen.physics@gmail.com}
\author{Z. R. Yang$^{1,7,8}$}
\email{zryang@issp.ac.cn}
\author{H. D. Zhou$^{9,10}$}
\email{hzhou10@utk.edu}
\author{X. F. Sun$^{2,7,8}$}
\email{xfsun@ustc.edu.cn}

\affiliation{$^1$Anhui Province Key Laboratory of Condensed Matter Physics at Extreme Conditions,
High Magnetic Field Laboratory, Chinese Academy of Sciences, Hefei, Anhui 230031, People's Republic of China}
\affiliation{$^2$Department of Physics, Hefei National Laboratory for Physical Sciences at Microscale, and Key Laboratory of Strongly-Coupled Quantum Matter Physics (CAS), University of Science and Technology of China, Hefei, Anhui 230026, People's Republic of China}
\affiliation{$^3$Fujian Institute of Research on the Structure of Matter, Chinese Academy of Sciences, Fuzhou, Fujian 350002, People's Republic of China}
\affiliation{$^4$National High Magnetic Field Laboratory,
Florida State University, Tallahassee, FL 32306-4005, USA}
\affiliation{$^5$Department of Physics, Florida State University,
Tallahassee, FL 32306-3016, USA}
\affiliation{$^6$State Key Laboratory of Surface Physics,
Center for Field Theory \& Particle Physics
Department of Physics, Fudan University, Shanghai, 200433, China}
\affiliation{$^7$Collaborative Innovation Center of Advanced Microstructures,
Nanjing, Jiangsu 210093, People's Republic of China}
\affiliation{$^8$Institute of Physical Science and Information Technology, Anhui University, Hefei, Anhui 230601, People's Republic of China}
\affiliation{$^9$Key laboratory of Artificial Structures and Quantum Control (Ministry of Education),
School of Physics and Astronomy, Shanghai JiaoTong University, Shanghai, 200240, China}
\affiliation{$^{10}$Department of Physics and Astronomy, University of Tennessee,
Knoxville, Tennessee 37996-1200, USA}
\date{\today}

\date{\today}

\maketitle

\subsection{Experimental Details}
The single crystalline samples of ZnCr$_2$Se$_4$ were grown by chemical vapor transport method using CrCl$_3$ as the transport agent in an evacuated sealed quartz tube~\cite{crystal}. During the growth process, the temperatures at two ends of the quartz tube were kept at 850 $^\circ$C and 950 $^\circ$C, respectively. The crystal quality was investigated by high-resolution synchrotron x-ray diffraction (Fig. 1) performed at 16-BM-D, HPCAT of Advanced Photon Source of Argonne National Laboratory with $\lambda$ = 0.4246 {\AA}. The Rietveld refinement of this room temperature pattern shows a lattice parameter $a$ = 10.491(8) {\AA} with stoichiometry as Zn$_{1.011(4)}$Cr$_{2.001(3)}$Se$_{3.987(2)}$. Within the errors of the experiment, the as grown crystals were stoichiometric with negilable site disorder.This result is consistent with the reported synchrotron x-ray and neutron powder diffraction studies on ZnCr$_2$Se$_4$\cite{ZnSeS5}. The dc magnetic properties were studied by
a superconducting quantum interference device (SQUID) magnetometer.
The ac susceptibility was measured at the National High Magnetic
Field Laboratory using the conventional mutual inductance
technique~\cite{AC}. The specific heat was measured in a
physical property measurement system (Quantum Design PPMS).
The thermal conductivity was measured by using a ``one heater,
two thermometers" technique~\cite{heater}. In all these
measurements, if the field is applied, the field is
along the [111] axis.

\begin{figure}[tbp]
\linespread{1}
\par
\begin{center}
\includegraphics[width=3.2 in]{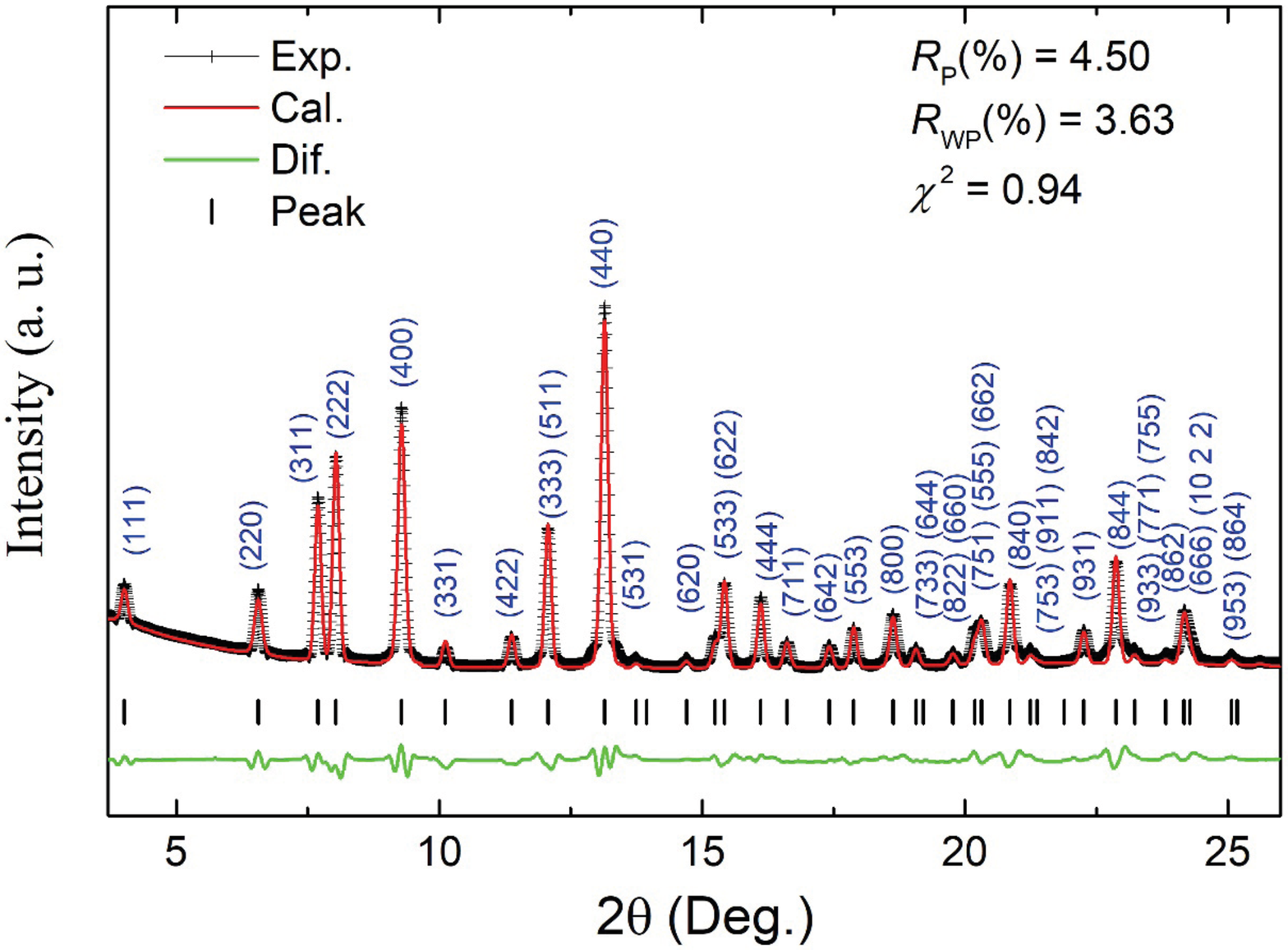}
\end{center}
\par
\caption{(Color online) The room temperature powder synchrotron x-ray diffraction spectrum of ZnCr$_2$Se$_4$. }
\label{Fig. S1}
\end{figure}